\documentclass[12pt]{article}
\usepackage{a4wide,epsfig}
\usepackage{epsfig}
\usepackage{rotating,amsmath,amssymb}
\usepackage{xspace}
\usepackage{longtable}
\usepackage{graphicx}

\newcommand{\bea}{\begin{eqnarray}}
\newcommand{\eea}{\end{eqnarray}}
\newcommand{\be}{\begin{equation}}
\newcommand{\ee}{\end{equation}}

\newcommand{\nue}{\ensuremath{\nu_{e}}\xspace}
\def\nubare{\ensuremath{\overline{\nu}_{e}}\xspace}

\newcommand{\numu}{\ensuremath{\nu_{\mu}}\xspace}

\newcommand{\dmtt}{\ensuremath{\Delta m^2_{23} \xspace}}

\newcommand{\thetaot}{\ensuremath{\theta_{13}}\xspace}

\newcommand{\numunue}{\ensuremath{\nu_\mu \rightarrow \nu_e}\xspace}

\newcommand{\pnumunue}{\ensuremath{P(\nu_\mu \rightarrow \nu_e)}\xspace}

\newcommand{\sigdm}{\ensuremath{{\rm sign}(\Delta m^2_{23})}\xspace}

\newcommand{\delCP}{\ensuremath{\delta_{\rm CP}}\xspace}

\newcommand{\stheta}{\ensuremath{\sin^2{2\theta_{13}}\xspace}}

\def\lsim{\mathrel{\rlap{\lower4pt\hbox{\hskip1pt$\sim$}}
    \raise1pt\hbox{$<$}}}                
\def\gsim{\mathrel{\rlap{\lower4pt\hbox{\hskip1pt$\sim$}}
    \raise1pt\hbox{$>$}}}                

\begin{document}

\renewcommand{\thefootnote}{\alph{footnote}}
  
\begin{center}
{\large \bf
 Next Challenge in Neutrino Physics: the \boldmath{\thetaot} Angle}
\vskip12pt
{M. Mezzetto }\\

{\it INFN - Sezione di Padova\\
 {\rm E-mail: mezzetto@pd.infn.it}}
\end{center}

\begin{abstract}A new generation of oscillation experiments
 optimized to measure \thetaot is ready to start.
 Performances, complementarity and competition of these accelerator and
reactor  experiments will be shortly illustrated.
The capability of measuring \thetaot with other neutrino sources, like solar, atmospheric,
supernovae neutrinos or neutrinos from a tritium source will be also discussed.
\end{abstract}
   
\section{Introduction}
 Three parameters of neutrino oscillations are still unknown:            
 the mixing angle \thetaot, the mass hierarchy \sigdm\ and the CP phase \delCP;
 they are all  fundamental parameters of the standard model.

 The mixing angle \thetaot is the key parameter of three-neutrino oscillations and regulates at the first order all the oscillation processes that could contribute to the measurement of \sigdm\ and \delCP. 

The best direct experimental limit on \thetaot comes from the Chooz reactor experiment~\cite{Chooz}.
A world limit can be derived~\cite{Schwetz:2008er} by a full $3\nu$ analysis of all the neutrino oscillation experiments,
see Tab.~\ref{tab:th13}.
The fact that the world limit provides a looser value than the Chooz limit indicates that
the best fit for \thetaot is different from zero,
although at small statistical significance, as discussed in \cite{Fit th13}.

\begin{table}[h]
\caption{The 90\%($3\sigma$) bounds (1 dof) on $\sin^2\theta_{13}$ from
an analysis of different sets of data  \protect\cite{Schwetz:2008er} }
{$
  \sin^2\theta_{13} \le \left\lbrace \begin{array}{l@{\qquad}l}
      0.060~(0.089) & \rm{(solar+KamLAND)} \\
      0.027~(0.058) & \rm{(Chooz+atm+K2K+MINOS)} \\
      0.035~(0.056) & \rm{(global\; data)}
    \end{array} \right.
$}
\label{tab:th13}
\end{table}

A preliminary analysis of the MINOS experiment \cite{Minos-nue} shows a $1.5 \sigma$ excess of \nue-like
events in the far detector, that could be interpreted as a manifestation of a non-zero value of \thetaot.

In the following will be reviewed the experimental potential of measuring \thetaot by using tritium sources, atmospheric neutrinos, supernova neutrinos and solar neutrinos. Then will be described the sensitivities of the next generation of accelerator and reactor neutrino experiments ready to start: T2K, NO$\nu$A, Double Chooz and Daya Bay. The complementarity of these measurements and the competition of the experimental sensitivities along the time will also be discussed.
\section{Tritium Experiments}
\begin{figure}[h]
    \includegraphics[width=0.37\textwidth]{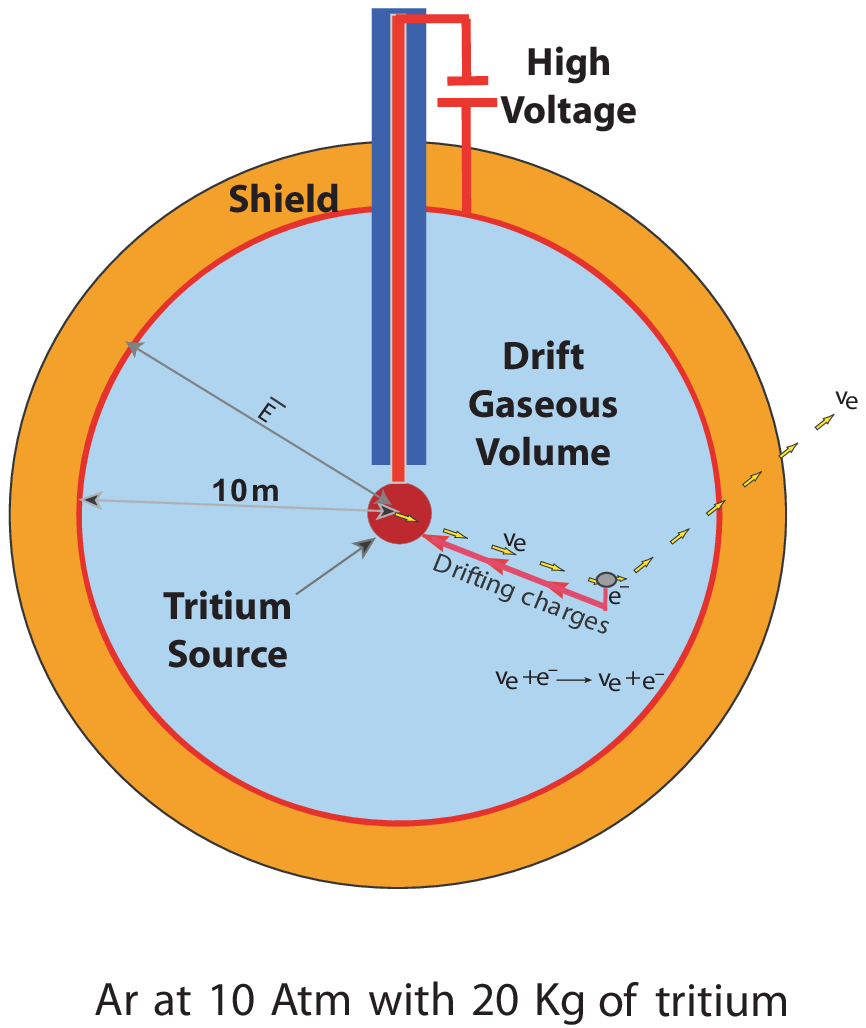}
    {\includegraphics[width=0.60\textwidth]{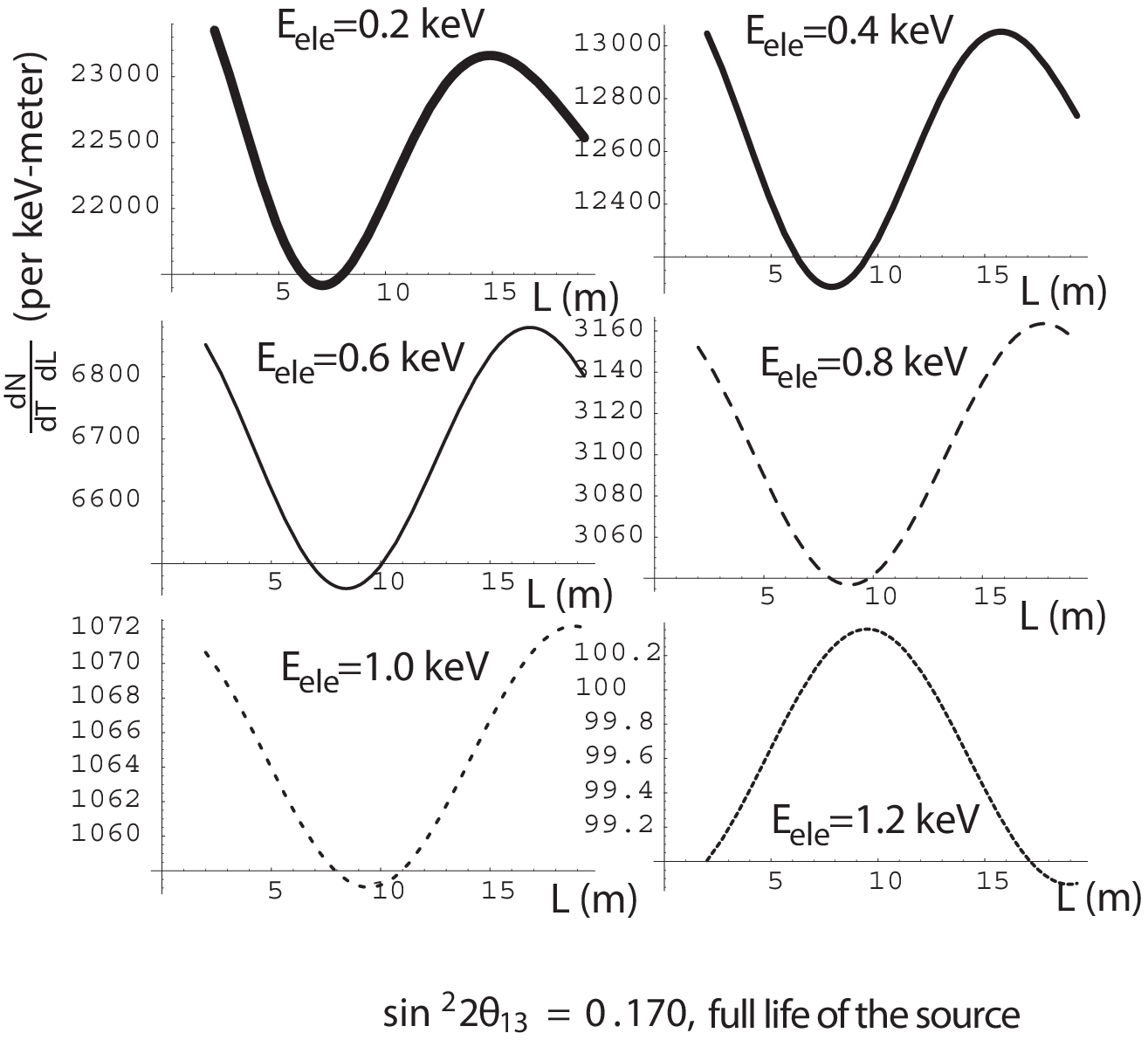}}
\caption{Left panel: sketch of the NOSTOS experiment. Right panel
The differential rate dN/dTdL (per keV-meter) for Ar at 10 Atm with 20 Kg
of tritium as a function of the source-detector distance (in m ), averaged over the
neutrino energy, for electron energies from top to bottom and left to right 0.2, 0.4,
0.6, 0.8, 1.0 and 1.2 keV. The results shown correspond to $\stheta = 0.170$. This
rate must be multiplied by $1−e^{−t/\tau}$ to get the number of events after running time
t. From~\protect\cite{Giomataris}.}
\label{fig:NOSTOS-results}
\end{figure}

Tritium has been considered as a possible source of neutrinos for table-top-like \nue disappearance experiments thanks to its small end-point energy: 18.6 KeV,
corresponding to a maximum baseline of 9.2 m for $\dmtt=2.5\cdot10^{-3}$ eV$^2$.
Giomataris et al. \cite{Giomataris} proposed to use an extremely intense source of tritium (200 MCi) surrounded by a spherical high-pressure gas TPC, 10 m radius, filled with argon at 10 atm and
 read-out by large surface Micromegas \cite{Micromega}, the NOSTOS experiment, Fig.~\ref{fig:NOSTOS-results} left.

In such configuration \nue disappearance would be measured as a function of the baseline, with some sensitivity to \thetaot.
Both elastic scattering and neutral current events would be detected in the TPC.
 These processes have different cross-section values as function of
the neutrino energy, so the path length disappearance shapes are different at different energies. As an example the path length curves as computed for \thetaot around the Chooz limit ($\stheta=0.170$) for the full life of the source ($T_{1/2}\simeq12.33$ yr) are displayed \cite{Giomataris}.
The potential of this setup cannot reach sensitivities much below the Chooz limit.

\begin{figure}
    {\includegraphics[width=\textwidth]{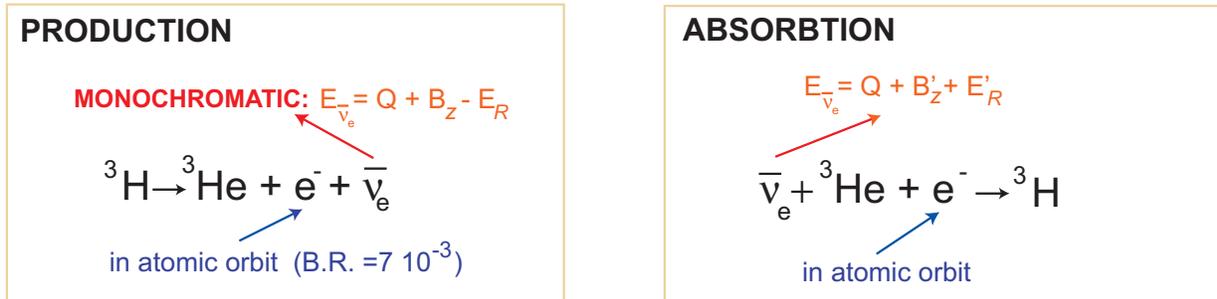}}
\caption{Sketch of the reaction scheme in a $^3$H-$^3$He experiment.}
\label{fig:mossbauer-scheme}
\end{figure}

A renewed interest in Tritium experiments came following the publication by Raghavan
\cite{Raghavan} about the possibility that
mono energetic antineutrinos emitted in the bound state beta-decay of $^3$H can be resonantly captured in $^3$He.
The reaction scheme is illustrated in Fig.~\ref{fig:mossbauer-scheme}.
The resonant character of the reactions is partially destroyed by the nuclear recoil energies  $E_R$, $E'_R$, nevertheless about 6 orders of magnitude could be gained with
respect to conventional neutrino cross sections.
The enhancement of the cross section could be up to  eleven orders of magnitude
 by embedding both $^3$H and $^3$He into solids  by which the broadening of the beam
due to nuclear recoil is severely suppressed by a mechanism similar to the M\"{o}ssbauer effect.

Under these assumptions an experiment exploiting 1 MCi
$^3$H metallic source and  100g  $^3$He
metallic detector could register $\sim 10^6$ events/day at 10 m allowing for
precision measurements of \thetaot, order of $\stheta\simeq 0.004\;(2\sigma)$ \cite{Minakata},
or for new ways of measuring the mass hierarchy \cite{Parke}.

Some questions anyway have been raised about the real possibility of gaining
these 11 orders of magnitude in cross section \cite{Potzel}.
The main question is how far can be set the same binding energies $B_z$, $B'_z$.
$^3$H and $^3$He atoms have different sizes, modifying the lattice structure and
so the binding energy.
This and other solid state effects can weaken the resonant peak,
 loosing up to 6 orders of magnitude in the cross section.

Furthermore the way itself in which the $^3$He lattice is produced: 
by  loading at first the lattice  with $^3$H and
waiting a long enough time to have it decayed, 
makes problematic a precise measurement
 of the  $^3$H generated by neutrino interactions.

While M\"{o}ssbauer neutrinos could be a very interesting setup to measure
neutrino oscillations, it appears that some
R\&D is needed to set the feasibility and the sensitivity of this experimental approach.
\section{Atmospheric Neutrinos}
\begin{figure}
    \centering
    {\includegraphics[width=0.4\textwidth]{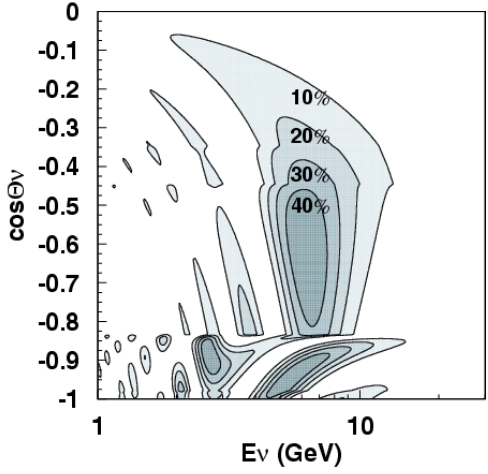}}
\caption{
Oscillation probability \pnumunue as function of the neutrino energy and
the zenith angle ($\cos{\Omega_\nu}=-1,\,0$ correspond
 to vertically upward and horizontal directions, respectively).
The three high probability ( $\geq 40\%$) regions are shown which correspond
to the MSW resonance at 3 GeV in the core layer, the
MSW resonance at 7 GeV in the mantle layer, and the enhancement
due to the core-mantle transition interference at the
energy between the two MSW regions. From~\protect\cite{SK-I atmo}.
}
\label{fig:sk-atmo-prob}
\end{figure}

The Super Kamiokande analysis of atmospheric neutrinos is sensitive to \thetaot through MSW transitions
 in the Earth, that can generate large
oscillation amplitudes, Fig.~\ref{fig:sk-atmo-prob}.

\begin{figure}
    {\includegraphics[width=\textwidth]{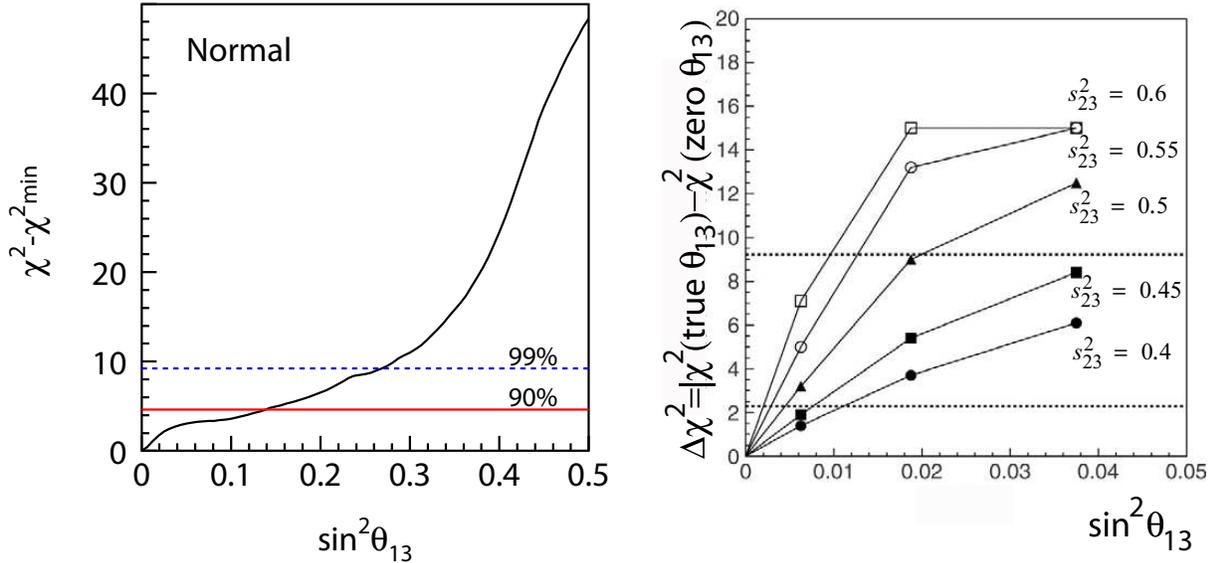}}
\caption{
$\chi^2-\chi^2_{\rm min}$ values of the fit to atmospheric neutrinos
in the Super Kamiokande experiment (left panel) \protect\cite{SK-I atmo},
 as a function of $\stheta$, assuming normal hierarchy.
Right panel: the same quantity extrapolated to longer exposures, from~\protect\cite{Kajita},
the two horizontal dashed lines are $3\sigma$ sensitivities of Super Kamiokande
for 20 and 80 years of data taking.
}
\label{fig:sk-atmo}
\end{figure}

The collaboration published limits about \thetaot based on a
three-neutrino analysis of atmospheric neutrino oscillations in the SK-I data taking \cite{SK-I atmo}, Fig.~\ref{fig:sk-atmo} left.

 While these limits are not competitive with the Chooz limit, 
the statistics so far collected is about two times bigger than what published.
Having a look to the
predicted sensitivities of Super Kamiokande as function of the exposure, see Fig.~\ref{fig:sk-atmo} right,
one can reasonably expect that a 3 $\nu$ analysis of the whole data set collected so far could allow Super Kamiokande
to reach a sensitivity equal to, or even better, than the Chooz limit.

This could be, in the short term, the best opportunity to see progress in the \thetaot hunt.

\section{Supernova Neutrinos}
Neutrinos generated by a supernova explosion can provide information about the \thetaot value, as discussed  in~\cite{th13-sn}. 

The main mechanism trough which neutrino rates at Earth are modulated by \thetaot is the MSW crossing probability at the high resonance region inside the supernova, that, after some approximations
\footnote{among which the non inclusion of collective neutrino effects\cite{Pantaleone}}, can be written as:

\begin{equation}
P_H \simeq \exp\bigg\{-\frac{\pi}{12}\bigg[\frac{10^{10}
MeV}{E}\bigg(\frac{\sin^32\theta_{13}}{\cos^22\theta_{13}}\bigg)
 \bigg(\frac{|\Delta m^2_{32}|}{1 eV^2}\bigg)C^{1/2}\bigg]^{2/3}\bigg\}.
\end{equation}

where the $C$ parameter takes into account the amount of electron capture during the star collapse, it is estimated
to be within the $[1,15]$ interval.

Several other supernova parameters influence the $\nu$ fluxes
like the $\nu$ flavour temperatures $T_\alpha$ and the pinching parameters (deviations from thermal energy distributions) $\eta_\alpha$.


To extract information about \thetaot, the experiments should provide spectral information about the different neutrino flavours with a sufficient statistics.
One could reasonably ask if the present generation of supernova neutrino detectors
has enough sensitivity to extract information about \thetaot in case of a 
supernova explosion.

\begin{figure}
\includegraphics[width=\textwidth]{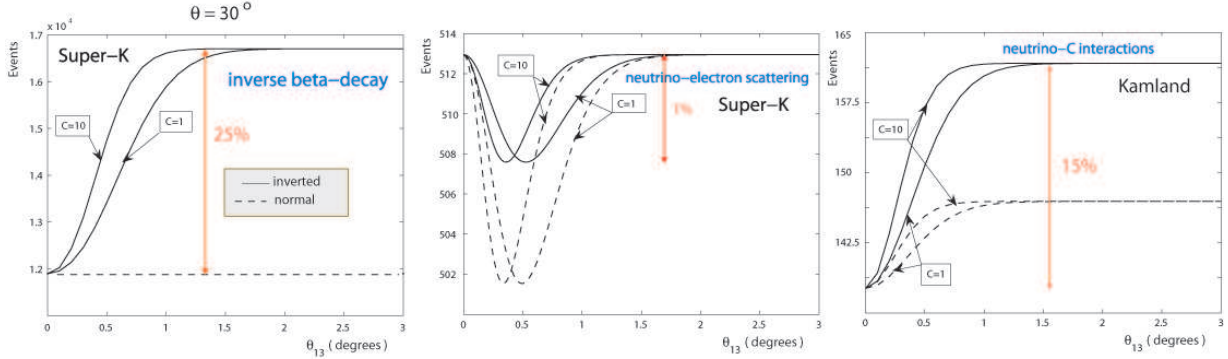}
\caption{
Estimated  supernova event number observed in three different experimental channels in  as a function of $\thetaot$
 computed for an incident angle $\theta = 30^\circ$ and a supernova explosion at 10 kpc.
 The solid curves correspond to the normal hierarchy,
and the dashed curves correspond to the inverted hierarchy. Also shown the variation of the rates by
changing the $C$ parameter, see the text, from 1 to 10. Elaborated from~\protect\cite{china-sn}.
}
\label{fig:SN-events}
\end{figure}

In Fig.~\ref{fig:SN-events} are displayed the estimated number of events detected by a supernova explosion at 10 kpc by inverse beta-decay in Super Kamiokande, neutrino-electron scattering in Super Kamiokande and neutrino-carbon interactions in KamLAND, as computed in \cite{china-sn}.
Due to the different contributions of neutrino flavors to the detected processes and the different cross-sections,
the detected rates have  different dependencies from \thetaot, ranging from a 1\% variation from small  ($\thetaot <1^\circ$ corresponding to $\stheta<10^{-3}$) to large values of \thetaot in case of the neutrino-electron scattering in Super Kamiokande to a 25 \% variation in case of inverse beta-decay in Super Kamiokande. 

It is certainly very difficult to make a detailed prediction of the capability of measuring \thetaot from supernova data, given the large number of supernova parameters to be fitted and the lack of detailed information about the efficiencies and the capability of correctly identify the interaction channels of the different detectors.

It seems anyway plausible that in case of a supernova explosion at 10 kpc the combination of the detected signals in the several running detectors, including also the spectral information, can allow to decide if \thetaot is bigger or smaller than about $1^\circ (\stheta=10^{-3})$,
 a very precious information, since next generation experiments optimized to measure \thetaot, see the following, will reach a sensitivity one order of magnitude smaller: $\stheta\simeq 10^{-2}$.

\section{Solar Neutrinos}
The non-zero value of \thetaot coming from  the world fits \cite{Fit th13} is driven by the tension of the KamLAND and
SNO measurements of the solar parameters. One could wonder if an improvement of the SNO and KamLAND results could allow for more significant evidence of non-zero values of \thetaot. 
SNO has already published the whole data set \cite{SNO}, it is expected to perform a full 3$\nu$ analysis of the data together with a lower detection threshold \cite{McDonald}, while significant improvements in statistics in KamLAND will be very slow (the published data set covers the period March 2002 to May 2007 \cite{KamLAND}).

A breakthrough in this field could come in case of doping with gadolinium of the Super Kamiokande detector \cite{Vagins1}, that could transform SK in a $\sim 30$ kton neutrino reactor detector.
In this configuration it has been shown \cite{Vagins2} the potential for a spectacular improvement of the precision of the measurement of the solar parameters (mostly $\Delta m^2_{12}$).

\section{Accelerator Neutrinos}

When matter effects are not negligible, 
 the
transition probability $\nu_e \to \nu_\mu$ ($\bar \nu_e \to \bar \nu_\mu$) can
be written as~\cite{PilarNufact}:
\begin{equation}
  P^\pm (\nu_e \to \nu_\mu)=
   X_\pm \sin^2 (2 \theta_{13}) + Y_\pm \cos ( \theta_{13} ) \sin (2 \theta_{13} ) 
  \cos \left ( \pm \delta - \frac{\Delta m^2_{23} L }{4 E_\nu} \right ) + Z \, , 
\label{eq:Donini}
\end{equation}
where $\pm$ refers to neutrinos and antineutrinos, respectively and
$a[{\rm eV}^2]=\pm 2\sqrt{2}G_Fn_eE_\nu=7.6{\cdot} 10^{-5}\rho[g/cm^3]E_\nu[{\rm GeV}]$
is the electron density in the material crossed by neutrinos. 
The coefficients of the two equations are:
$
X_\pm = \sin^2 (\theta_{23} ) 
\left ( \frac{\Delta m^2_{23} }{ | a - \Delta m^2_{23}| } \right )^2 \sin^2 \left ( \frac{|a - \Delta m^2_{23}| L}{ 4 E_\nu } \right ) 
$, \\
$
Y_\pm = \sin ( 2 \theta_{12} ) \sin ( 2 \theta_{23} ) 
\left ( \frac{\Delta m^2_{12} }{ a } \right ) \left ( \frac{\Delta m^2_{23} }{ |a - \Delta m^2_{23}| } \right )
\sin \left ( \frac{ a L }{ 4 E_\nu } \right ) \sin \left ( \frac{ |a - \Delta m^2_{23}| L }{ 4 E_\nu } \right ) 
$,\\
$
Z = \cos^2 (\theta_{23} ) \sin^2 (2 \theta_{12}) 
\left ( \frac{\Delta m^2_{12} }{ a } \right )^2 \sin^2 \left ( \frac{a L }{ 4 E_\nu } \right )
$.

The \numunue\  transitions are dominated by the solar term, anyway, at the distance
 defined by the $\Delta m^2_{23}$  parameter, they are driven by the \thetaot term which is
 proportional to $\sin^2{2\thetaot}$.
 Moreover $P(\nu_\mu  \rightarrow \nu_e)$ could be strongly
 influenced by the unknown value of \delCP and \sigdm.

Given the complexity of the \numunue\  transition formula  it will be very
difficult for pioneering experiments to extract all
the unknown parameters unambiguously. Correlations are present between \thetaot and
\delCP \cite{PilarNufact}. Moreover, in absence of information about
 the sign of $\Delta m^2_{23}$~\cite{MinakataDege,Barger:2001yr} and the
approximate $[\theta_{23}, \pi/2 - \theta_{23}]$ symmetry for the
atmospheric angle~\cite{FogliDege}, additional clone solutions rise up. In general,
the measurement of $P(\nu_\mu \to \nu_e)$ and $P(\bar \nu_\mu \to \bar
\nu_e)$ will result in eight allowed regions of the parameter space,
the so-called eightfold-degeneracy~\cite{Barger:2001yr}.

Experimental \thetaot  searches at the accelerators
 look for  evidence
 of \nue\   appearance in an intense \numu beam
 in excess of what is expected from the solar terms.

 The $\nu_{\mu} \rightarrow \nu_e$ experimental sensitivity
 with conventional $\nu_\mu$ beams is limited by an
 intrinsic $\nue$ beam contamination of about 1\%.
   Furthermore, neutral pions in both neutral current and charged current interactions
  can fake an electron providing also a possible background for the $\nu_e$'s.

  Therefore the measurement of the $\theta_{13}$ mixing angle
 will require
  neutrino beams with high performances in terms of intensity, purity
  and associated systematic errors.
  Detectors should combine a very large mass
            with high granularity and resolution necessary to keep detector
            backgrounds at  as low as possible rates.
  Ancillary experiments to measure the meson production (for the
            neutrino beam knowledge), the  neutrino cross-sections, the
            particle identification capability will become necessary. 
            The Harp hadroproduction experiment at CERN PS \cite{Harp-Al,Harp-Be}
            for instance, measured the hadroproduction for the proton energy
            and target material of the
            K2K and MiniBooNE experiments, giving a fundamental contribution
            to the reduction of the systematic errors.
            The NA61 experiment at CERN~\cite{NA61} is going to measure the hadroproduction
            for the T2K setup.

In the following we will focus on T2K and NO$\nu$A, the approved
\thetaot optimized accelerator experiments.

There are several proposals for next generation experimental setups,
based on conventional neutrino beams, capable to significantly improve the
sensitivity of T2K and NO$\nu$A in the future \cite{Dusel,MODULAr,SPL-SB}.
 Ultimate performances
in neutrino oscillation searches at the accelerators can be
reached by neutrino beams based upon innovative concepts, like neutrino
factories \cite{Nufact} and beta beams \cite{beta beams}.

\subsection{T2K}
\begin{figure}
\centering
\includegraphics[width=0.6\textwidth]{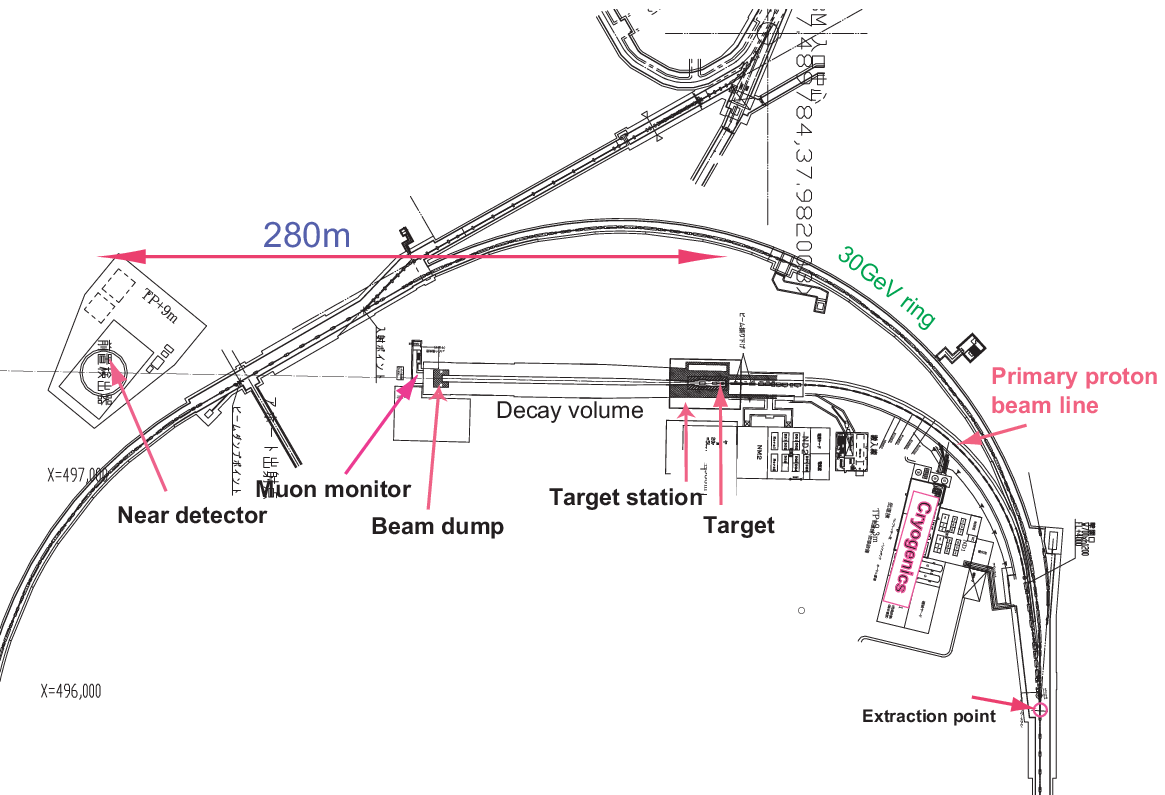}
\includegraphics[width=0.38\textwidth]{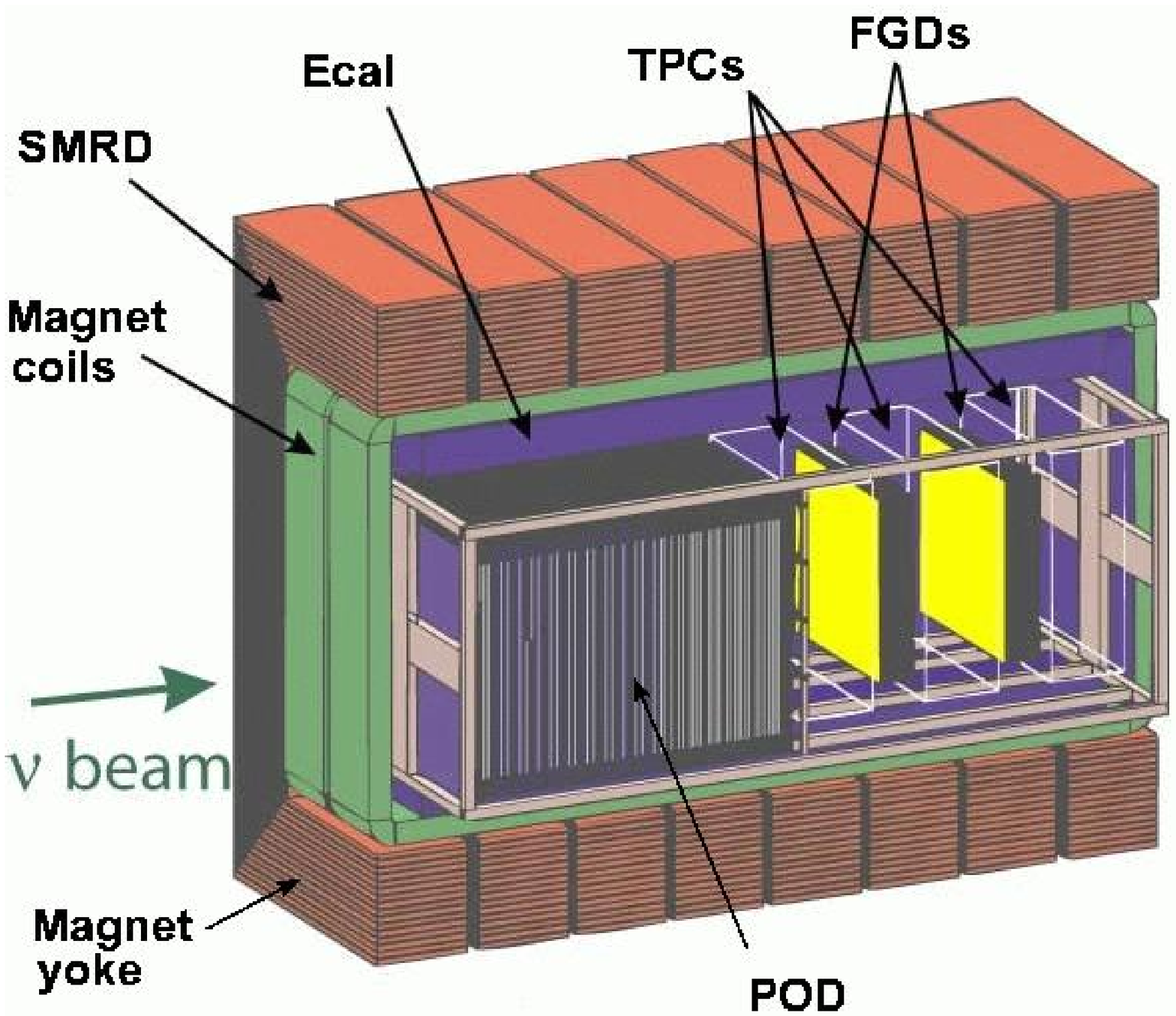}   
 \caption{Left panel: the layout of the T2K beam line, showing the location of primary proton beam line,
target station, decay volume, beam dump, muon monitors and near neutrino detectors. 
Right panel: sketch of the T2K ND280 near detector.}
\label{fig:t2k-beamline}
\end{figure}

The T2K (Tokai--to--Kamioka) experiment~\cite{T2K}   will use a 
high intensity  off--axis neutrino beam generated by a 30 GeV  proton beam at
  J-PARC 
(Japan Proton Accelerator Research Complex) fired to the  Super Kamiokande   
detector, located 295 km from the proton beam target. 
The schematic view of the T2K neutrino beam line is shown in Fig.~\ref{fig:t2k-beamline} left.

A sophisticated near detector complex (ND280)  will 
be built at a distance of 280 m from the target. 
This complex has two detectors: one on-axis (neutrino beam
monitor)    and  the other off--axis.
This off-axis detector (Fig.~\ref{fig:t2k-beamline} right)
is a spectrometer built inside
 the magnet of the former experiments UA1 and Nomad,
 operating with a magnetic field of 0.2 T.
It includes a Pi-Zero detector (POD), a tracking detector made by three  time projection 
chambers (TPC's) and two fine grained scintillator detectors (FGD's), a $4\pi$
electromagnetic calorimeter (Ecal), and a side muon range detector~(SMRD).   
Neutrino rates in the close detector will be about 160000 \numu (3200 \nue) interactions/ton/yr at the nominal
beam intensity of 0.75 MW$\cdot 10^7$ s. 

 ND280 is expected to calibrate
 the absolute energy scale of the neutrino spectrum  
with 2\% precision, measure the non-QE/QE ratio at the 5-10\% and
 monitor the neutrino flux with better than 5\% accuracy.
 The  momentum resolution of muons from the 
charged current quasi-elastic interactions~(CCQE) should be better than 10\%. 
The $\nu_e$ fraction should be measured with an uncertainty better than 10\%.
A measurement of  the neutrino  beam direction, with a precision better 
than 1 mrad, is required from the on-axis detector.

\begin{figure}
\includegraphics[width=0.35\textwidth]{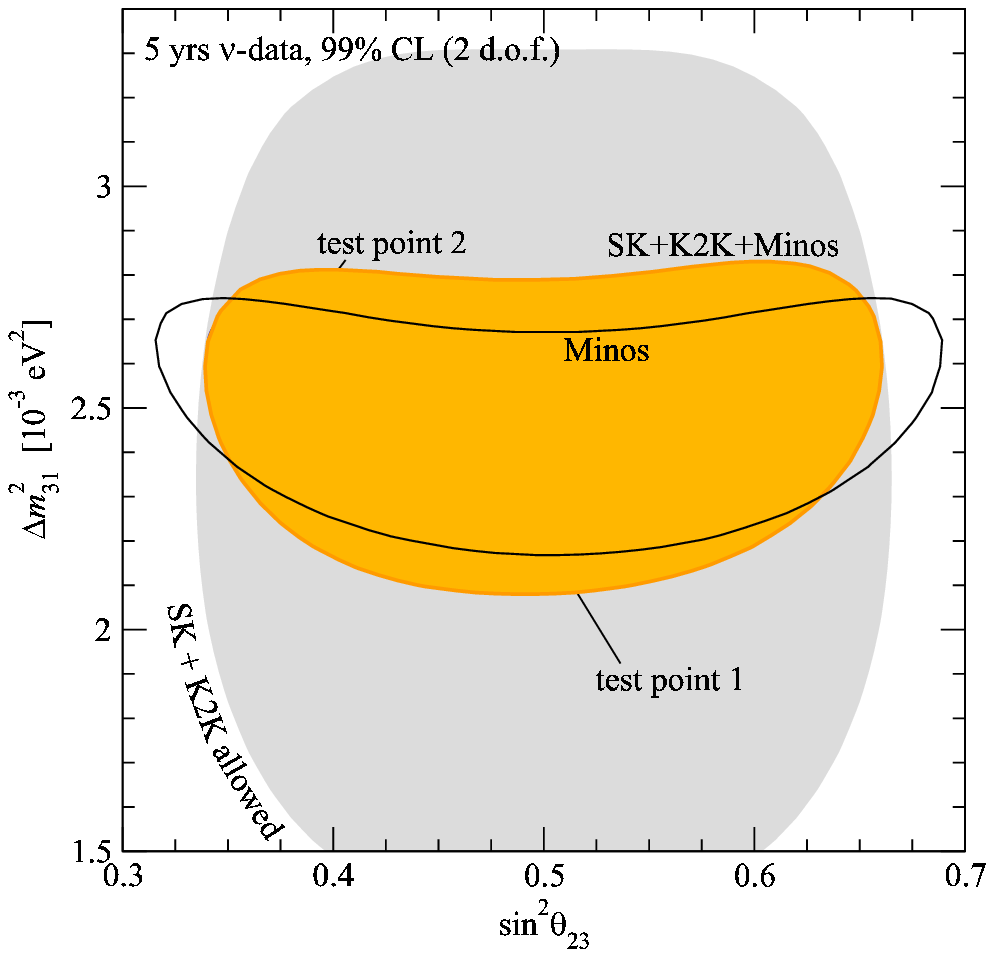}   
\includegraphics[width=0.63\textwidth]{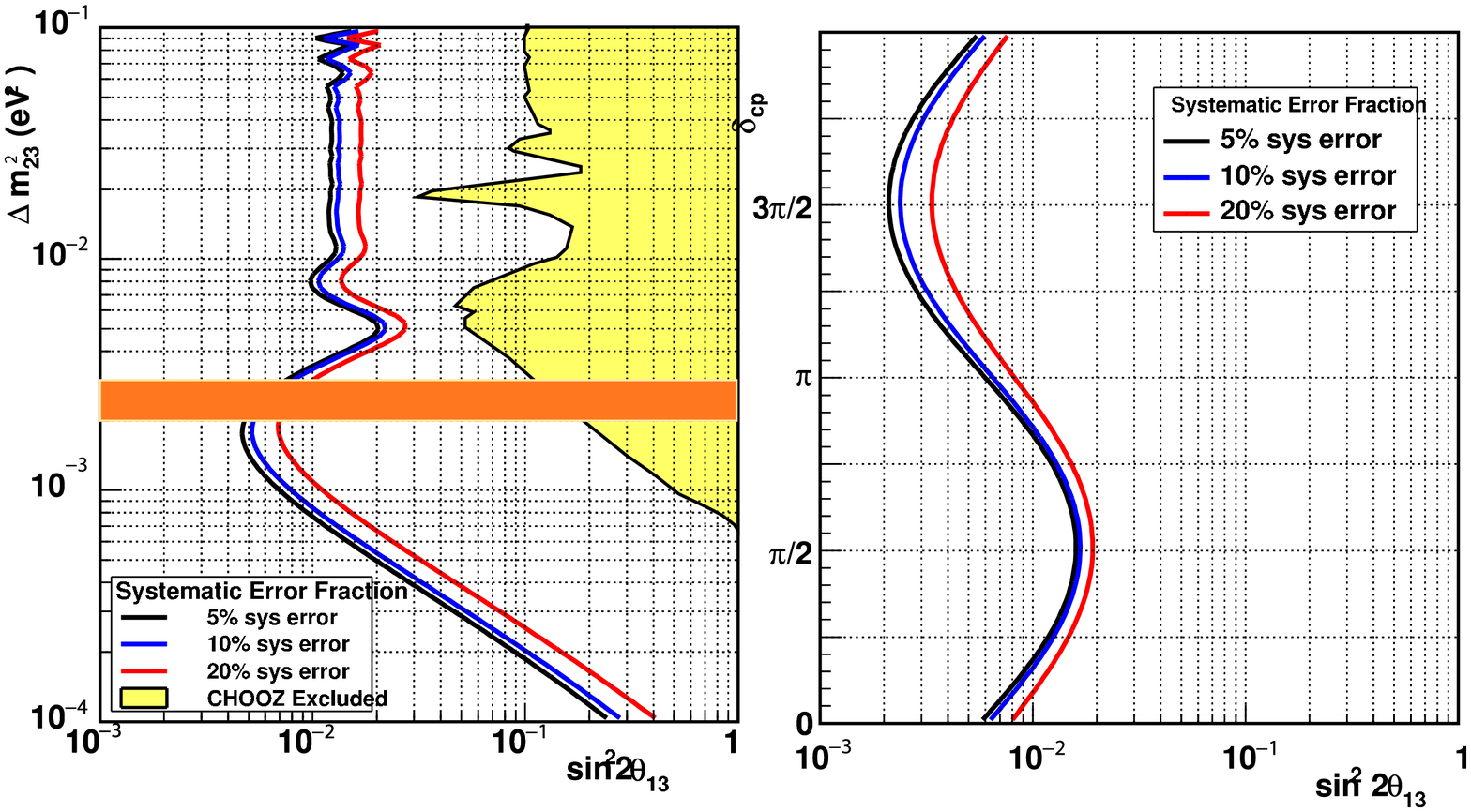}   
\caption{Left panel: 99\% CL contours for two test points selected within the 99\% allowed values by
the world fits \protect\cite{Schwetz:2008er}. They are computed for 5 years data taking at 
0.75 MW/year, and 5\% systematic errors.  The T2K values are taken by \protect\cite{Campagne07}.
Also shown are the allowed regions by fits to SuperKamiokande+K2K and to Minos only.
Central panel: 90\% CL sensitivity to $\stheta$, 
 computed for 5 years data taking at 0.75 MW/year. 
compared with the Chooz limit in the $\stheta$ vs$\dmtt$ plane,
assuming $\delCP=0$ and normal hierarchy, for three different choices of the systematic errors.
Right plot: the same sensitivity computed in the $\delCP$ vs $\stheta$ plane,
assuming $\dmtt=2.5\cdot 10^{-3}$ eV$^2$ and normal hierarchy.}
\label{fig:T2K-th13}
\end{figure}

The sensitivity of T2K
in measuring the atmospheric parameters through the \numu disappearance is shown in Fig.~\ref{fig:T2K-th13} (T2K is expected to collect about 16000 \numu interactions in 5 years
at the nominal beam intensity, neglecting the oscillations).

Fig.~\ref{fig:T2K-th13} center and right show 
the sensitivity in measuring \thetaot. The experiment will reach
a factor 20 improvement with respect to the Chooz limit.

The commissioning of the neutrino beam line successfully started on April, 24, 2009. Data taking is 
scheduled to start end 2009, integrating the first year 0.1 MW$\cdot 10^7$ s protons, allowing for a $\stheta$
sensitivity of $\stheta \simeq 0.1$ (90\%CL, $\dmtt=2.5\cdot 10^{-3}$ eV$^2$,
$\delCP=0$, normal hierarchy.).

The T2K setup has been designed to be  scalable \cite{T2K}. The J-PARC beam intensity
can be upgraded up to 1.6 MW 
and a new  water \v{C}erenkov detector with a fiducial 25 times bigger than
Super Kamiokande, Hyper Kamiokande~\cite{T2K}, can be build in the Kamioka region.

\subsection{NO$\nu$A}

 The NO$\nu$A experiment~\cite{Nova}  will run at an upgraded NuMI neutrino beam
 ($6.5 {\cdot} 10^{20}$ pot/year,
corresponding to a beam power of 700 kW;
$E_{\nu} \sim 2 $ GeV and a $\nu_e$ contamination less than $0.5 \%$)
 at baseline of
  810 Km, 14 mrad off-axis. 
 The start-up phase of the experiment is funded and the fully approval is expected
within 2009.
 The far detector will be a 15 kt ``totally active'' tracking liquid scintillator,
 scheduled to be fully operational by the end of 2013.
 The close detector will be a 215 ton replica of the far detector, placed 14 mrad off the
 NuMI beam axis at a distance of 1 km from the target.
  NO$\nu$A plans to run 3 years in neutrino mode and 3 years in antineutrino mode.
Since NO$\nu$A
will reach similar \thetaot sensitivities of T2K with several years of
delay, cfr.~Fig.~\ref{fig:th13vstime}, the focus of the experiment is to provide data on the neutrino
mass hierarchy, where NO$\nu$A has a clear advantage with respect to T2K
thanks to the longer baseline. These searches require                    
a statistically significant antineutrino run.
In doing that 
 NO$\nu$A can also provide first indications about the
 range of \delCP and informations about \thetaot complementary to T2K.

  As a second phase,   the  NuMI beam
  intensity could be increased to 1.2 MW (``SNuMI'') or to
2.3 MW (``Project X'') in case  the new proton driver of
 8 GeV/c and 2 MW will be built at FNAL.

\section{Reactor Experiments}
Reactor experiments can measure \thetaot by detecting \nubare
 disappearance at the atmospheric $\Delta m^2$. The oscillation disappearance $P_{\bar{\nu}_{e}\bar{\nu}_{e}}$ can be expressed as:
\begin{equation}
1-P_{\bar{\nu}_{e}\bar{\nu}_{e}} \, \simeq \, \sin^22\theta_{13} \, \sin^2(\Delta m^2_{31}L/4E) \, + \,
(\Delta m^2_{21}/\Delta m^2_{31})^2) \, (\Delta m^2_{31} L/4E)^2  \cos^4\theta_{13} \, \sin^22\theta_{12}~
\label{eq:Pee}
\end{equation}
showing a direct connection between $P_{\bar{e}\bar{e}}$ and \thetaot, with no 
interference by \delCP and \sigdm \footnote{
this also means that in the long term there is no way
to directly measure leptonic CP violation with a reactor experiment}.
\begin{figure}
    \centering
    \includegraphics[width=0.5\textwidth]{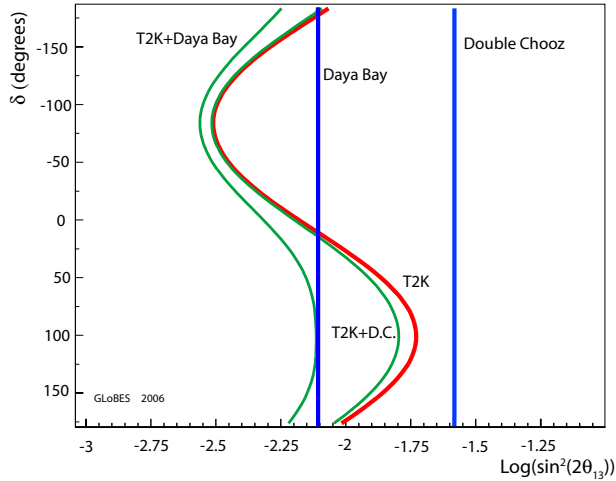}
\caption{90\% CL sensitivity of T2K, from~\protect\cite{T2K}, Double Chooz,from~\protect\cite{Double Chooz} , and 
Daya Bay, from~\protect\cite{Daya Bay}. Also shown are the combinations of T2K with Double Chooz and with Daya Bay, as computed with Globes \protect\cite{Globes}.}
\label{fig:th13-combined}
\end{figure}

The deep difference between the appearance formula Eq.~\ref{eq:Donini}
 and the disappearance Eq.~\ref{eq:Pee} suggests that 
the two experimental approaches are truly complementary.
This is illustrated in Fig.~\ref{fig:th13-combined} where the nominal, final, sensitivity of T2K is compared with the sensitivities of
the reactor experiments Double Chooz and Daya Bay.
 While the appearance sensitivity is modulated by the unknown value of \delCP, the disappearance sensitivity is flat. Their combination provides a powerful sensitivity plot where the \delCP modulation is reduced and the overall sensitivity increased
\footnote{on the other hand reactor experiments, having no sensitivity to the atmospheric parameters, need the information of an accelerator experiment to delimit the \dmtt\ range where they probe \thetaot}.

\begin{figure}
    \centering
    \includegraphics[width=0.48\textwidth]{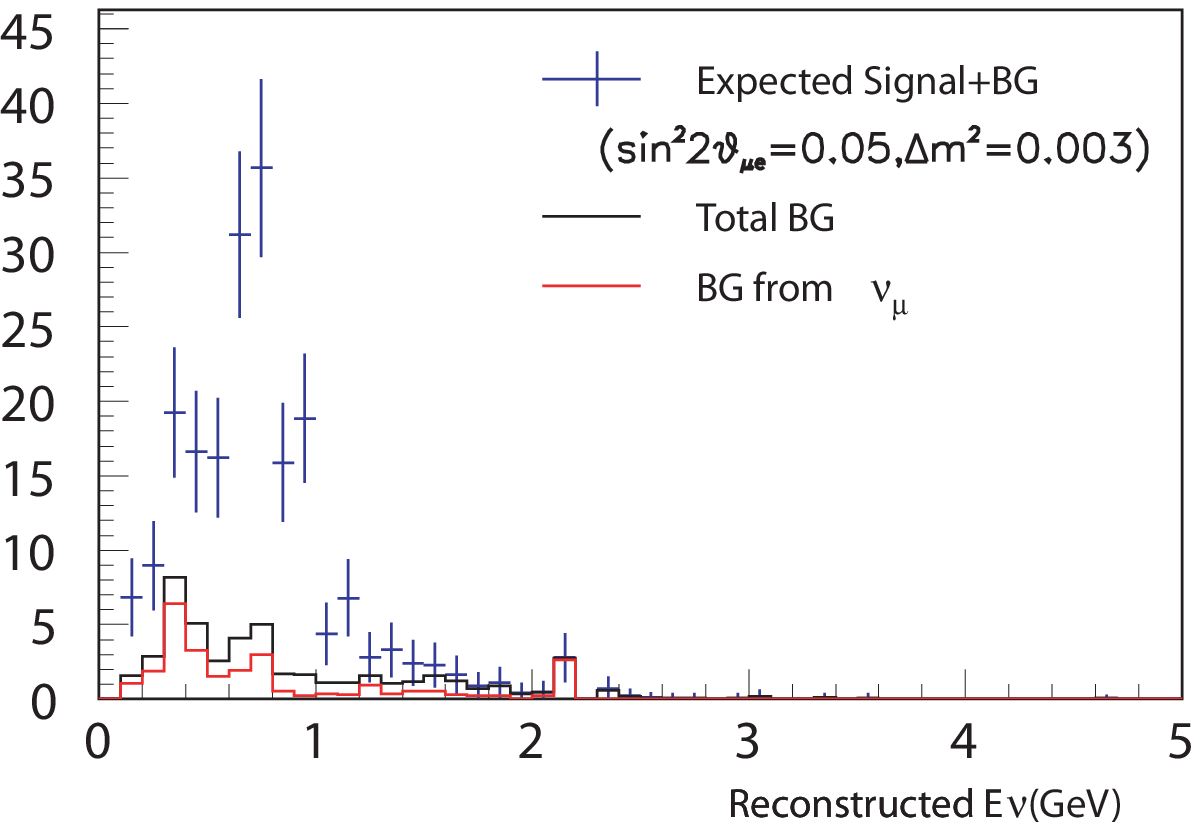}
    \includegraphics[width=0.48\textwidth]{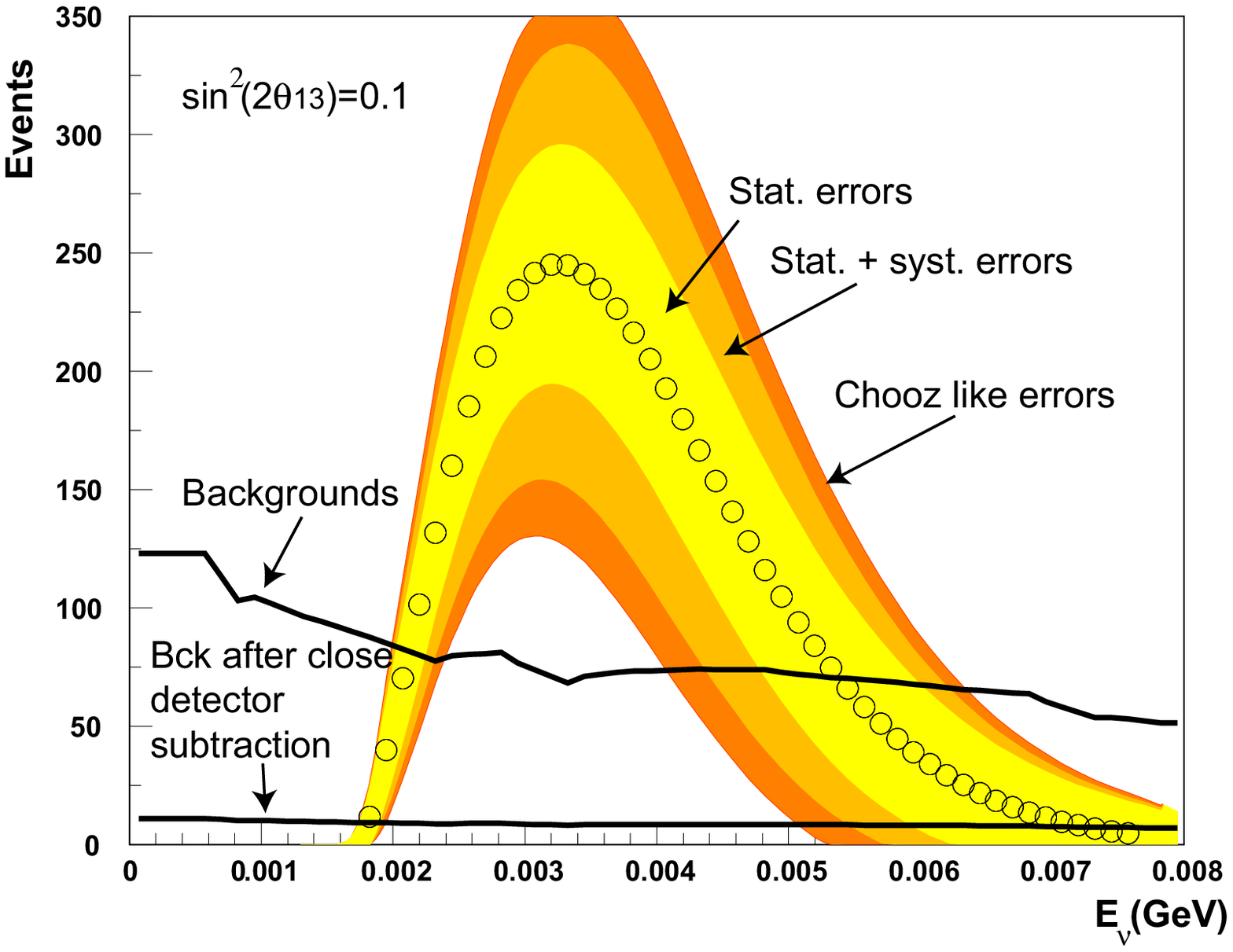}
\caption{Left panel: signal and backgrounds events for T2K, computed for $\stheta=0$, $\delCP=0$, $\Delta m^2_{23}=2.5\cdot 10^{-3}$ eV$^2$ and normal hierarchy, from~\protect\cite{T2K}.
Right panel: number of disappeared events in Double Chooz under the same conditions.
Also shown statistic and expected systematic errors, together with systematic errors 
as big as the former Chooz experiment. The signal is compared with the expected background
 rate, before and after the subtraction of the close detector data. Elaborated from~\protect\cite{Double Chooz}.}
\label{fig:reactor-signal}
\end{figure}

Appearance experiments are limited by statistics and background rates, while reactor experiments are limited by systematic errors, as illustrated by Fig.\ref{fig:reactor-signal} where are compared the signal distributions for T2K and Double Chooz computed for $\stheta=0.1$.

%
\subsection{Double Chooz}
The Double Chooz experiment \cite{Double Chooz}
will be installed near the Chooz two-core (4.27+4.27 GW) nuclear power plant. The far detector, a 8.3 t gadolinium loaded liquid scintillator detector, will be placed in the existing site of the previous Chooz experiment, 1.05 km from the reactor cores, at a depth of about 300 m.w.e.
The close detector, identical to the far detector, will be placed at about 400 m from the reactor cores (not at the exact relative distance of the far detector), at a depth of 115 m.w.e.
The experiment aims to an overall systematic error of 0.6\%, the far detector is expected to begin data taking end of 2009, while the close detector should be put in operation by end of 2011.
The \thetaot sensitivity of the experiment as function of time is shown in Fig.~\ref{fig:dChooz-sens} left.

\subsection{Daya Bay}
The Daya Bay experiment \cite{Daya Bay} will receive neutrino by two nuclear plants: 
Daya Bay and LingAo located in the south of China.
The two nuclear plants are about 1100 m apart.
 Each nuclear plant has two cores running.
 Another two cores, called LingAo II, are expected to be commissioned by the end of 2010.
The thermal power of each core is 2.9 GW,
hence the existing total thermal power is 11.6 GW, and will be 17.4 GW after 2010.
The basic experimental layout of Daya Bay
consists of three underground experimental halls, one far and two near, linked by horizontal tunnels.
Each near hall will host two 20 t gadolinium doped liquid scintillator detectors, while the
far hall will host four such detectors.

\begin{figure}
  \centering
    \includegraphics[width=0.90\textwidth]{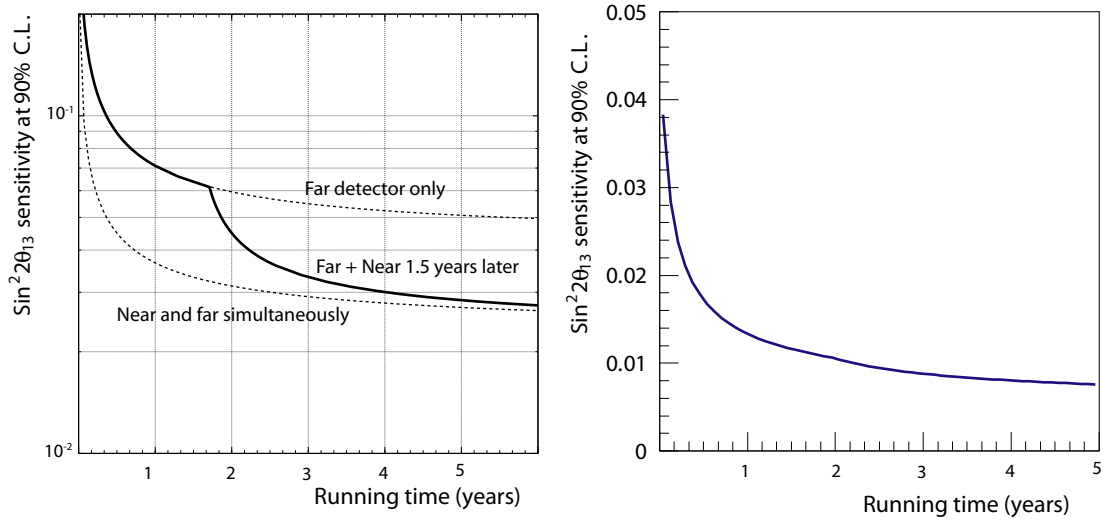}
\caption{Left panel:
 expected $\stheta$ sensitivity, 90\%CL, of the Double Chooz experiment, computed for $\dmtt=2.5\cdot10^{-3}$
eV$^2$,
year zero is expected to be end 2009, from~\protect\cite{Double Chooz}.
Right panel: the same for the Daya Bay experiment, from~\protect\cite{Daya Bay}, where year zero is expected to be summer 2011.}
\label{fig:dChooz-sens}
\end{figure}

The experiment aims to an overall systematic error of 0.38\% (for a comparison of the systematic errors of Double Chooz and Daya Bay see~\cite{Mention}), the far detectors are
 expected to begin data taking mid of 2011. 
The \thetaot sensitivity of the experiment as function of time is shown in Fig.~\ref{fig:dChooz-sens} right.

\section{Guessing the Future}
\begin{figure}
\centering
{\includegraphics[width=0.8\textwidth]{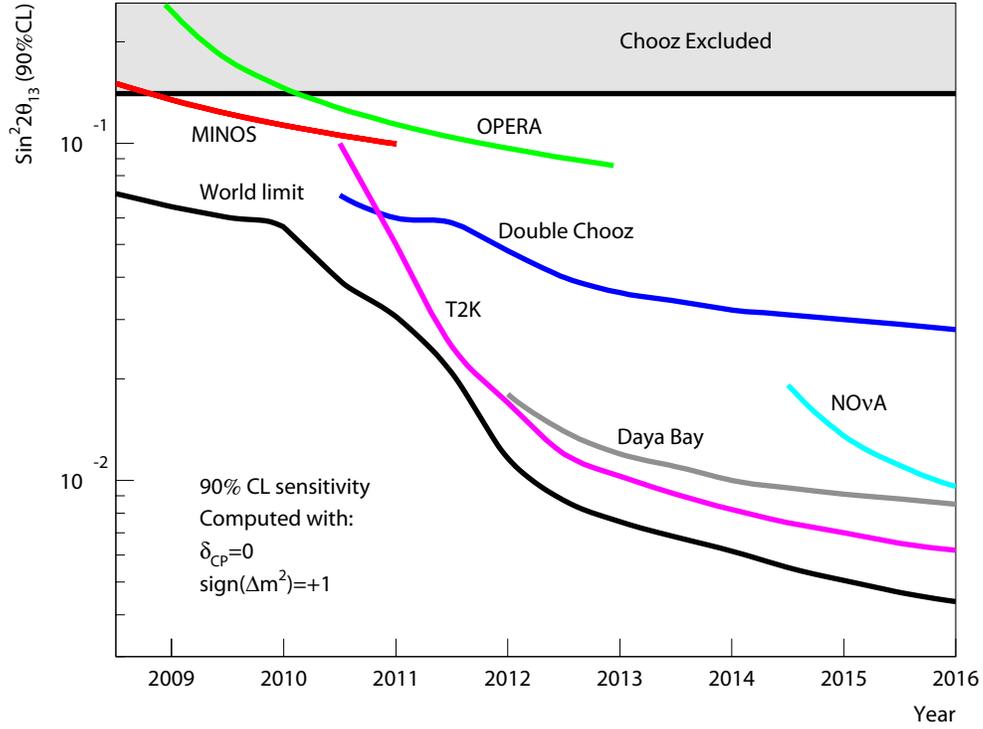}}
    \caption{Evolution of experimental $\sin^2{2\thetaot}$ sensitivities 
    as function of time.  
    All the sensitivities are taken from the proposals of the experiments.
    For T2K it is assumed a beam power of 0.1 MW the first year, 0.75 MW from 
    the third year and a linear transition in between.
    NO$\nu$A sensitivity is computed for $6.5\cdot 10^{20}$ pot/yr, 15 kton detector
    mass, neutrino run.
    Accelerator experiments sensitivities are computed for $\delCP=0$ and normal
    hierarchy, for all the experiments $\dmtt=2.5\cdot10^{-3}$ eV$^2$.
    The sensitivity curves are drawn starting after six months of data taking.
    }
    \label{fig:th13vstime}
\end{figure}
It is of some interest to have a look to the expected sensitivities
of accelerator and reactor experiments in the near future.
Fig.~\ref{fig:th13vstime} shows the evolution of the \thetaot sensitivities
 as a function of the time.
From the plot one can derive that in the next 5 years or so the 
\thetaot parameter will be probed with a sensitivity about 25 times better than
the present limit.

\begin{figure}
\centering
{\includegraphics[width=0.7\textwidth]{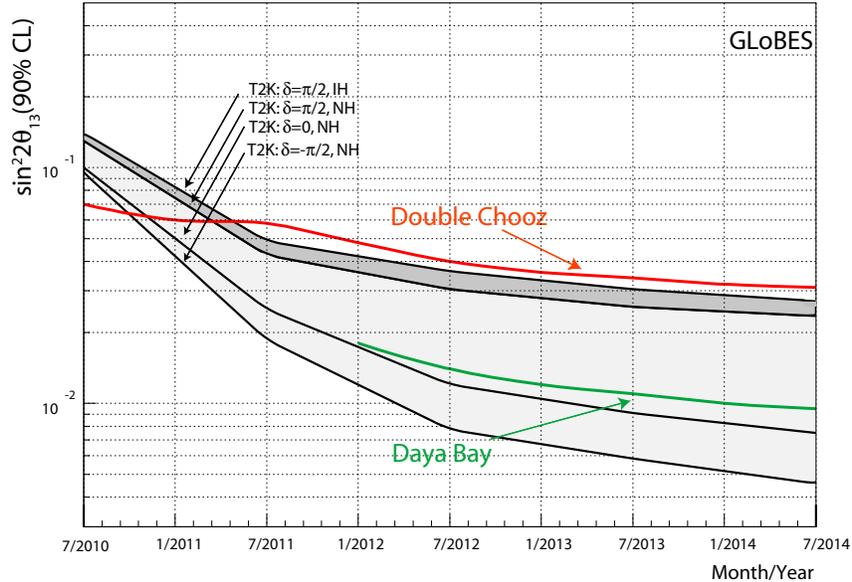}}
    \caption{Evolution of experimental $\sin^2{2\thetaot}$ sensitivities 
    of T2K and the reactor experiments Double Chooz and Daya Bay
    as function of time, under the same assumptions of Fig.~\protect\ref{fig:th13vstime}.
    The T2K sensitivity is computed with Globes~\protect\cite{Globes},
    using the Globes library,
    and shown as a band of values computed for
    different values of $\delCP$ and for normal (NH) and 
    inverted (IH) hierarchy.}
    \label{fig:time-evolution}
\end{figure}

Since the
T2K \thetaot sensitivity depends from the unknown \delCP parameter
and from the choice of the mass hierarchy (cfr. Fig.~\ref{fig:T2K-th13} right and
\ref{fig:th13-combined}), a more detailed comparison of the time evolution
of the T2K and reactors sensitivities should take into account that
T2K will provide a band of excluded values of \thetaot and not just a single value.
This is shown in Fig.~\ref{fig:time-evolution}.

From the plot some considerations can be taken:
\begin{itemize}
	\item
Double Chooz is very competitive in the first years of operation, when the information
of the close detector probably will not be available.
	\item
The time evolution of beam power of T2K is crucial. It is impossible to state now which
time evolution will have the J-PARC neutrino beam line, based on a totally new accelerator
complex, so the sensitivity shown here is just a personal educated guess.
	\item
Also for Daya Bay the schedule is critical. Very important will be also the goal of
very small systematic errors claimed by the experiment.
\end{itemize}

This discussion is based on sensitivities, where no signal in the detectors is assumed.
In case of \thetaot in the reach of those experiments, their information will be
truly complementary to measure the true value of the parameter, for a discussion 
under this hypothesis see for instance \cite{Lindner}.

\end{document}